\documentclass[aps,showpacs,prl,twocolumn,superscriptaddress]{revtex4}
\usepackage{graphicx}% Include figure files
\usepackage{amssymb}
\usepackage{amsmath}
\usepackage{psfrag}%To include LaTeX subscripts on figures
\usepackage{color}
\newcommand{\derx}{\partial_x}

\begin{document}

\title{
Bloch oscillations in one-dimensional spinor gas}
 \author{D.M.~Gangardt}
\affiliation
{School of Physics and Astronomy, University of Birmingham, Edgbaston,
Birmingham, B15 2TT, UK  \email[e-mail: ]{d.m.gangardt@bham.ac.uk}
}
%\email[e-mail: ]{d.m.gangardt@bham.ac.uk}
\author{A. Kamenev }
 \affiliation{
School of Physics and Astronomy, University of Minnesota,
   Minneapolis, MN 55455 }
%\email[e-mail: ]{kamenev@physics.umn.edu}

\date{\today}


\begin{abstract}
A force applied to a spin-flipped particle in a one-dimensional spinor gas may
lead to Bloch oscillations of particle's position and velocity.  The existence
of Bloch oscillations crucially depends on the viscous friction force exerted
by the rest of the gas on the spin excitation.  We evaluate the friction in
terms of the quantum fluid parameters. In particular, we show that the
friction is absent for integrable cases, such as $SU(2)$ symmetric gas of
bosons or fermions.  For small deviations from the exact integrability the
friction is very weak, opening the possibility to observe Bloch oscillations.
\end{abstract}

\maketitle

%\section{Introduction}
%\label{sec:intro}

Dynamics of ultracold atomic gases with internal (spinor) degree of freedom
has been a focus of a number of recent experiments
\cite{Lewandowski_2002}.
The observed collective phenomena have revived interest in earlier theoretical
works \cite{Bashkin_1981} on spin waves in helium and opened a
possibility to study nonequilibrium dynamics of quantum liquids.

Due to an unprecedented degree of experimental control it is possible to
excite a few atoms into a different hyperfine internal  
state \cite{Chikkatur_00}.
This leads to effective spin excitations which may be regarded as
impurities moving through the quantum liquid formed by the majority spins.  A
similar setup was investigated in the context of He$^3$ and He$^4$ mixtures
\cite{LandauKhalatnikov1949ViscosityI,BaymEbner1967Phonon}.
It was realized that an external particle is ''dressed''
to form a collective excitation which energy--momentum relation at small
momenta $P$ is quadratic $\varepsilon(P) \approx \mu_d + P^2/2M^*$. The
correlations manifest themselves in quasiparticle effective mass $M^*$ being
different from the bare mass $M$, as well as in the friction exerted on the
quasiparticle by the rest of the liquid.

\begin{figure}[t]
  \centering
 \psfrag{0}{0}
 \psfrag{p}{$P$}\psfrag{E}{$\varepsilon(P)$}\psfrag{pin}{$\pi \hbar n$}
 \psfrag{2pin}{$2\pi \hbar n$}\psfrag{-pin}{-$\pi \hbar n$}
 \psfrag{pq}{$P+q$}
  \includegraphics[width=8cm]{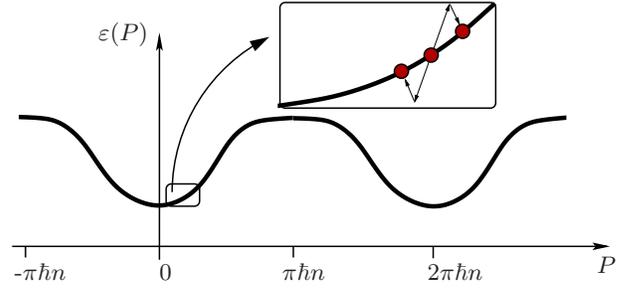}
  \caption{Dispersion relation  of a spin excitation  in
    1D Bose liquid. Inset:  two-phonon processes
    leading to a dissipation. The arrows, with the slope  given by 
    the sound velocity $c$, represent absorption and emission of
    long wavelength phonons. }
  \label{fig:disp}
\end{figure}

The collective nature of the excitations is especially apparent in
one-dimensional (1D) systems where the strong effects of interactions
beyond mean field were recently observed in experiments with cold 
atoms \cite{Tonks_Experiments}. 
In addition to the strong mass
renormalization \cite{Fuchs_2005}, power law  behavior of responce functions
\cite{zvonarev_2007,Matveev08,Kamenev08}, the dispersion relation of the
excitations $\varepsilon (P)$ was shown  \cite{Matveev08,Lamacraft08} to be 
strongly modified by the interactions: parabolic at small $P$, 
it is actually a {\em periodic} function of the momentum with the
period $2\pi \hbar n$, 
%i.e.  $\varepsilon(P)=\varepsilon(P+2\pi\hbar n)$, 
see Fig.~\ref{fig:disp}. Here $n$ is a 1D density of the gas and periodicity 
stems from the fact that total momentum $P_\mathrm{tot} = 2\pi \hbar n$ can be
transferred to the gas as a whole at no energy cost in thermodynamic limit.  
% Here $n=N/L$ is a density of 1D gas, $L$ is the length
% and $N$ is number of particles.  The periodicity stems from the fact that
% momentum $P_\mathrm{tot} = N\times2\pi\hbar/L$,
% may be transferred to the gas as a whole, while
% the corresponding excitation energy $P_\mathrm{tot}^2/(2Nm)$ vanishes in the
% thermodynamic limit. 
The periodicity of the dispersion relation drastically
affects the dynamics of spin excitations under an influence of the external
gravitational force $F$, which becomes uncompensated if 
the hyperfine state of impurity atoms is insensitive to the magnetic 
field of the trap \cite{Donner_2007}.
Indeed,  momentum of impurity evolves
according to $\dot P=F$ and its velocity 
$V=\partial\varepsilon(P)/\partial P$ is
a periodic function of momentum and thus exhibits {\em Bloch oscillations}.
Bloch oscillations are usually associated with an accelerated quantum particle
in presence of a static periodic potential, see, e.g., Ref.~\cite{Morsch_2001}
for a recent experimental realization.  However, the existence of Bloch
oscillations in 1D spinor condensates does {\em not} rely on a presence of an
external periodic potential.  It is the 1D quantum liquid itself that provides
a quasi-periodic potential with the lattice spacing $n^{-1}$ and thus $2\pi
\hbar n$ reciprocal vector.

Contrary to a static periodic potential, a 1D quantum liquid exhibits quantum
and thermal fluctuations.  Because of the latter the spin excitations with
periodic dispersion relation are also subject to a dissipation.  A possibility
to observe Bloch oscillations depends crucially on the strength of such a
dissipation, i.e. the friction force exerted on the spin-flipped particle by
the quantum liquid. Indeed, in a presence of friction the equations of motion
for the spin excitation take the form
\begin{equation}\label{EofM} \dot P=F-\kappa V\,; \quad \quad V=\dot
X= \partial \varepsilon/\partial P ,
\end{equation} where $\kappa=\kappa(V)$ is the coefficient of viscous friction
\cite{foot1}.  Consider, e.g., the strong coupling limit where
$\varepsilon(P) \approx \mu_d +(2\hbar^2 n^2/M^*) \sin^2(P/2\hbar n)$
\cite{Matveev08} and $\kappa\approx \mbox{const}$.  Integrating
Eqs.~(\ref{EofM}), one finds for the  drift velocity:
\begin{equation}\label{driftV}
  V_{\rm drift} = \kappa^{-1}
  \left\{ \begin{array}{ll} F\,; & \quad F < F_c\,, \\
      F-\sqrt{F^2-F_c^2}\,; &
      \quad F > F_c\,, \end{array} \right.
\end{equation}
where the critical force is $F_c=\kappa \hbar n/M^*$.  As a result the drift
velocity is a non-monotonic  function of the applied force and friction
$\kappa$. As discussed below, the latter depends strongly on the temperature
and interaction parameters, leading to a non-monotonic  dependence of $V_{\rm
drift}$ on them too. In the strong forcing regime, $F>F_c$, the drift motion
is superimposed with the Bloch oscillations with the period $\Delta T =2\pi
\hbar n/\sqrt{F^2-F_c^2}$.  Notice that both the period and the amplitude
of the Bloch oscillations  diverge as $F$ approaches $F_c$ frome above.

To find whether the Bloch oscillations regime is experimentally accessible,
one needs to have a detailed understanding of the friction $\kappa(V)$. The
friction originates from the interactions between the spin-excitation and the
density excitations (phonons) of the rest of the gas.  Since the latter
propagate with the speed of sound $c$ typically much larger than velocity of
the particle $V$, the single-phonon processes do not lead to dissipation
according to the Landau criterion \cite{LandauKhalatnikov1949ViscosityI}.
However, the two-phonon processes (see inset in Fig.~\ref{fig:disp}) do lead
to energy and momentum relaxation which scales as $T^7$ in 3D
\cite{LandauKhalatnikov1949ViscosityI,BaymEbner1967Phonon} and as $T^4$ in 1D
\cite{castro96} (at small temperature and $V<c$). In this paper we evaluate
the magnitude of $T^4$ friction and express it in terms of ``Bose liquid''
parameters.

Dimensional arguments \cite{castro96} lead to $\kappa \sim T^4/(c^4 \hbar^3
n^2)$, which for typical experimental parameters would make the critical force
$F_c$ rather large (see below). Fortunately the actual friction may be {\em
parametrically} smaller than this estimate. The reason is that for certain
sets of parameters the corresponding 1D models are \emph{integrable}. One
example is provided by $SU(2)$ symmetric spinor liquid of bosons or fermions
with point-like interactions \cite{CN_Yang_1967}, where the bare masses and
interaction constants of both spin species are the same: $M=m$ and $G=g$
\cite{foot4}.  The other example is the Tonks gas $g\to \infty$ with $M=m$ and
an arbitrary interaction constant $G$ between the spin excitation and the
liquid \cite{McGuire_1965,Castella_Zotos_1993}.  As we explicitly verify here
in such integrable cases the dissipation is absent \cite{Zotos02}  
and the spin excitations
have infinite life-time even at finite temperature. Thus in the vicinity of
integrable points the dissipation is strongly suppressed (see, e.g.,
Ref.~\cite{Newton_craddle} for experimental evidence in the spinless case).

In a generic case we found that the friction coefficient may be expressed
through the thermodynamic properties of the gas.  To this end one needs to
know the chemical potentials of the majority $\mu(n)$ and minority $\mu_d(n)$
spin species as functions of majority spins concentration $n$ (the minority
concentration is vanishingly small). For the friction coefficient at small
velocity, $V\ll c$, we found
\begin{eqnarray} \label{rof0}
\kappa &=& \frac{16\pi^3}{15}\, \frac{T^4}{c^4 \hbar^3 n^2}\, \Bigg[
1-\delta_l-\frac{M}{m} \\ &+& \frac{m}{M^*}
\left(1-\delta_l-\frac{M}{m}\right)^2 + \frac{\alpha
n^2}{mc^2}\left(1-\delta_l -\frac{\alpha_d}{\alpha} \right)\Bigg]^2 \nonumber
\,,
\end{eqnarray} where the sound velocity is related to the compressibility as
$mc^2/n=\partial \mu/\partial n$ and $\alpha = \partial^2 \mu/\partial n^2$,
while
\begin{equation}\label{alphas}
  \frac{mc^2}{n}(1-\delta_l)= \frac{\partial
    \mu_d}{\partial n}\,;\quad\quad
  \alpha_d= \frac{\partial^2 \mu_d}{\partial n^2}\,.
\end{equation}
In the $SU(2)$ symmetric case the equality of bare masses and interaction
constants imply $\delta_l=0$ and $\alpha_d=\alpha$. As a result the friction
is absent $\kappa=0$. The same is true for the $M=m$ Tonks gas, which is
equivalent to an impurity moving through the non-interacting Fermi gas
\cite{foot2}. Finding the chemical potentials and the effective mass for
non-integrable cases is a difficult task. Analytic progress may be achieved in
two limits:

{\em Weak coupling limit:} $g\ll \hbar^2 n/m$ and $G\ll \hbar^2 n/M$.
According to Bogoliubov theory the mean-field equations of state are
$\mu = g n $ and $\mu_d = G n$, while $M^*\simeq M$ \cite{Fuchs_2005}.
Extracting parameters as in Eq.~(\ref{alphas}) for both spin
excitation and the liquid yields
\begin{eqnarray}
  \label{eq:kappa_weak}
   \kappa(0) =  \frac{16\pi^3}{15}\, \frac{T^4}{c^4\hbar^3 n^2}
   \left( \frac{G}{g}\right)^2 \left(\frac{m G}{M g } - 1 \right)^2\, .
\end{eqnarray}
Keeping the velocity dependence of the friction force, one finds
$\kappa(V)=\kappa(0)/(1-V^2/c^2)$. As a result, the drift velocity of the
spin excitations cannot surpass the speed of sound, $V_{\rm drift}\leq c$,
making it hard to observe Bloch oscillations in weakly interacting gases.

{\em Strong coupling limit:} (i) $M\neq m$ and $g,G\to \infty$. Both majority
and minority spins may be considered as two free Fermi gases separated by
impermeable mobile wall. Demanding the equal pressure of the two gases, one
may find their total kinetic energy and thus respective chemical potentials:
$\mu =(\pi\hbar n)^2/(2m)$ and $\mu_d = (m/M)^{1/3} \mu $. Substituting it in
Eqs.~(\ref{rof0}), (\ref{alphas}) and employing that in this regime the
effective mass is large $M^*/M\sim GM/\hbar^2 n$, one finds
\begin{eqnarray}
  \label{eq:kappa_strong}
\kappa = \frac{16\pi^3}{15} \frac{T^4}{c^4 \hbar^3 n^2}
\left[ \left(\frac{m}{M}\right)^{1/3} -\frac{M}{m}\right]^2\, .
\end{eqnarray}
(ii) $M= m$ and $g,G\gg \hbar^2 n/m$. In this case we treat the corresponding
Fermi gases as being weakly interacting and find leading perturbative
corrections to the chemical potentials: $\mu = (\pi^2\hbar^2 /2m)(n^2 -
16\hbar^2 n^3/3gm)$, while $\mu_d = (\pi^2\hbar^2/2m)(n^2-4\hbar^2 n^3[1/g +
1/3G]/m)$.  Inserting it in the Eq.~(\ref{rof0}), one finds that in this order
of expansion all the terms cancel each other. Such a cancelation is not
expected if the order $n^4$ is kept in the chemical potentials.  One can thus
estimate the friction coefficient as
\begin{eqnarray}
  \label{eq:kappa_strong1}
  \kappa \sim
  \frac{\pi^3 T^4}{c^4 \hbar^3 n^2} \left(\frac{\hbar^2 n}{gm}\right)^4 \left(
  1- \frac{g}{G}\right)^2\, .
\end{eqnarray}
Since the effective mass of the spin excitations in the strong coupling regime
is large $M^*\gg M$ \cite{Fuchs_2005}, its velocity is small $V\ll c$ and the
friction coefficient is practically velocity-independent.  It is in this
regime where the Bloch oscillations are most likely to be observed.

We turn now to the derivation of Eq.~(\ref{rof0}). If the temperature is less
than the chemical potential $T\ll \mu$ the {\em density} excitations of the
majority spin gas may be described  
\cite{Popov_Lifshits_Pitaevskii}
by the effective 1D hydrodynamic Hamiltonian
\begin{eqnarray}
  \label{eq:ham_eff} H_\mathrm{ph} = \int\!dx\left[ \frac{1}{2m} (n+\rho)
(\derx \vartheta)^2 + \frac{m c^2}{2\,n}\, \rho^2 + \frac{\alpha}{3!}\,\rho^3
\right],
\end{eqnarray}
hereafter $\hbar =1$.
Here $\rho(x)$ is the operator of density fluctuations on top of the uniform
density $n$ and its canonically conjugate phase operator $\vartheta(x)$ is
related to the superfluid velocity $v_{s} = \derx \vartheta/m$. The quadratic
(Luttinger liquid) part of this Hamiltonian describes phonons with the linear
dispersion relation $\omega(q)=c|q|$. To have a consistent description of the
interactions between phonons and the spin excitations, one needs to take into
account non-linear interactions of phonons between themselves
\cite{BaymEbner1967Phonon}. They are described by the terms $\sim \rho (\derx
\vartheta)^2$ and $\sim \rho^3$. The coefficients in front of them are
dictated by Galilean invariance for the former, and by expansion of
$\mu(n+\rho(x))$ up to the second order in $\rho$ for the latter.

The spin excitation may be thought of as a quantum particle described by the
canonically conjugated coordinate $X$ and momentum $P$. Its interactions with
the density fluctuations are encoded in the chemical potential
$\mu_d(n+\rho(X))\approx (mc^2/n)(1-\delta_l)\rho(X)+ (\alpha_d/2)\rho^2(X)$,
cf. Eq.~(\ref{alphas}).  The interactions of this particle with the superfluid
velocity $v_{s}(X)$ may be found \cite{BaymEbner1967Phonon} by noticing that
in the reference frame where $v_{s}=0$ the energy of the particle with
momentum $P$ is given by the dispersion relation $\varepsilon(P)$.  In the
laboratory frame momentum of such particle is $P+Mv_{s}$, while its energy is
$\varepsilon_{v_{s}}(P+Mv_{s})=\varepsilon(P)+Pv_{s}+Mv_{s}^2/2$.  Changing to
a momentum in the laboratory frame, and keeping terms up to the second power
in $v_{s}$, one finds $\varepsilon_{v_{s}}(P)=
\varepsilon(P)+(P-MV)v_{s}+(M^2\varepsilon''(P)-M)v_{s}^2/2$, where the
particle velocity is $V=\varepsilon'(P)$. At a sufficiently small velocity
$V=P/M^*$ and $\varepsilon''(0)=1/M^*$ one thus finds for the Hamiltonian of
the spin excitations
\begin{eqnarray}
  \label{eq:h_p} H_d = \frac{(P + \delta\! M v_{s})^2}{2M^*} - \frac{\delta\!
M v_{s}^2}{2} + \frac{mc^2}{n}(1-\delta_l)\rho+ \frac{\alpha_d}{2}\rho^2 ,
\end{eqnarray} where $\delta\!M=M^*-M$.
In the absence of interactions $\delta M = M^*-M=0 $ and
Hamiltonian (\ref{eq:h_p}) is independent of the fluid velocity.  The
terms containing $\rho=\rho(X)$ and $v_{s}=\partial_x\vartheta(X)/m$ can be
regarded as effective interaction potential dependent on the particle
coordinate $X$. This introduces a preferential frame for the moving particle.

It is convenient to perform canonical transformation of the particle momentum
$P+ \delta\!M v_{s} \to P $ along with the fluid ``coordinate''
$\rho(x)-\big(\delta\!M/m\big)\rho_d(x)\to \rho(x)$, where
$\rho_d(x)=\delta(x-X)$ is the density of the particle. 
The changes induced by this transformation to the fluid 
Hamiltonian, 
Eq.~(\ref{eq:ham_eff}), are absorbed in the modified impurity Hamiltonian 
\begin{equation}
  \label{eq:h_p1} H_d = \frac{P^2}{2M^*} + \frac{m c^2}{n} \left(1-\delta_l
+\frac{\delta\!M}{m} \right) \rho
+\frac{\alpha}{2}\left(\frac{\alpha_d}{\alpha}+\frac{\delta\!M}{m} \right)
\rho^2.
\end{equation} The second term here describes processes in which one phonon is
absorbed or emitted, while the third one is responsible for the two-phonon
processes.  Due to the quadratic dispersion relation of the spin excitation
and linear dispersion of phonons, the one-phonon processes do not lead to
dissipation, which is just another statement of Landau criterion.

\begin{figure}[t]
  \centering
 \psfrag{a}{a}\psfrag{b}{b}\psfrag{c}{c}\psfrag{d}{d}
 \psfrag{r}{$\rho$}\psfrag{t}{$\partial_x \vartheta$}
  \includegraphics[width=8cm]{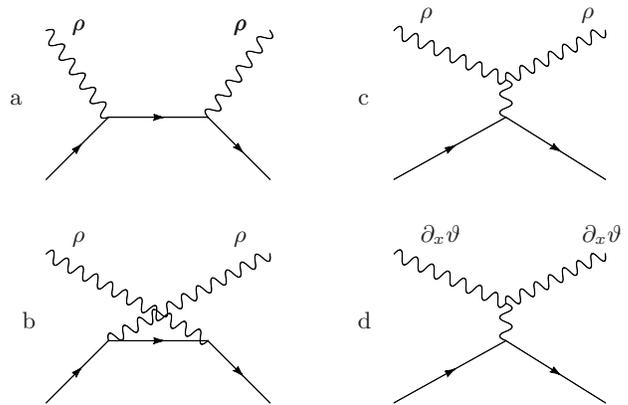}
  \caption{Second order diagrams contributing to 2-phonon amplitudes. Spin
  excitation is represented by a full line, while phonons by wavy lines.
  Diagrams a,b and c contribute to $\Gamma_\rho$, Eq. (\ref{eq:gamma-rho}), 
  while diagram d represents
  $\Gamma_\vartheta$, Eq.~(\ref{eq:gamma-theta}). }
  \label{fig2}
\end{figure}

We thus focus on the {\em two-phonon} amplitude.  The latter originates from
the last term in Eq.~(\ref{eq:h_p1}) as well as the following second order
processes: (i) second order in $\rho(X)=\int\! dx\, \rho(x)\rho_d(x)$
interaction vertex, Fig.~\ref{fig2}a,b; (ii) first order in $\rho\rho_d$ and
first order in $\rho^3$, Fig.~\ref{fig2}c or $\rho(\partial_x\vartheta)^2$
phonon nonlinearity vertexes, Fig.~\ref{fig2}d.  It is the destructive
interference of these second order processes which is responsible for the
partial or even complete suppression of the dissipation.  Evaluating the
corresponding diagrams according to the standard rules, one derives an
effective Hamiltonian of the spin excitation \cite{foot3}
 \begin{eqnarray}
  \label{h_eff} H_d^\mathrm{eff} =\frac{P^2}{2M^*} - \frac{1}{2}\,\Gamma_\rho
\big[\rho(X)\big]^2 - \frac{1}{2}\, \Gamma_\vartheta \big[\derx\vartheta
(X)\big]^2\,,
\end{eqnarray}
where the effective two-phonon amplitudes are given by
\begin{eqnarray}
                                                    \label{eq:gamma-rho}
  \Gamma_\rho\!\! &=&
\!\!
\frac{m^2c^2}{n^2M^*} \left(1-\delta_l +
\frac{\delta\!M}{m} \right)^2\! -
\alpha\left(\frac{\alpha_d}{\alpha}-1+\delta_l\right) \! ; \\
\Gamma_\vartheta\!\! &=&\!\! \frac{1}{m}\left(1-\delta_l +\frac{\delta\!M}{m}
\right) .
                                                    \label{eq:gamma-theta}
\end{eqnarray}

The momentum relaxation rate may be evaluated in the second order in the
two-phonon amplitudes $\Gamma_{\rho,\vartheta}$, averaged over the Luttinger
(quadratic) part of the phonon Hamiltonian (\ref{eq:ham_eff}). This does {\em
not} assume smallness of the amplitudes $\Gamma_{\rho,\vartheta}$, but rather
gives the leading low-temperature, $T\ll \mu$, contribution. The semiclassical
equation of motion for the spin excitation acquires a form
\begin{eqnarray}
  \label{eq:momentum_dot}
\dot P = -\frac{1}{4} \left(\Gamma_\rho-\frac{m^2c^2}{n^2}\,
\Gamma_\vartheta\right)^2 \int \frac{dq}{2\pi}\,\, q \,\Pi(q,qV)\,,
\end{eqnarray}
where $\Pi(q,\omega)$ is the imaginary part of the Fourier transform of
$\theta(t)\langle[\rho^2(x,t),\rho^2(0,0)]\rangle$ response function of
the phonon gas. The latter is given by
\begin{equation}
  \label{eq:Pi} \Pi (q,\omega) = \!\frac{n^2}{4m^2c^3}\!
\left(q^2\!-\!\frac{\omega^2}{c^2}\right) \!  \! \left(\!
\coth\frac{cq\!-\!\omega}{4T} - \coth\frac{cq\!+\!\omega}{4T}\! \right).
\end{equation}
For spin excitations of small velocity, $V\ll c$, one finds
\begin{equation}
  \label{eq:Pi1} \Pi (q,qV) =
\!\frac{n^2}{8m^2c^3}\,\frac{q^3}{T}\,\frac{V}{\sinh^{2}(cq/4T)}\,  .
\end{equation}
Substituting Eqs.~(\ref{eq:gamma-rho}), (\ref{eq:gamma-theta}) and
(\ref{eq:Pi1}) in
Eq.~(\ref{eq:momentum_dot}), one finds the friction force $\dot P=-\kappa V$
with the friction coefficient $\kappa$ given by Eq.~(\ref{rof0}).
Notice that in the integrable $SU(2)$ symmetric case the two-phonon amplitudes
$\Gamma_\rho=M^*c^2/n^2$ and $\Gamma_\vartheta=M^*/m^2$ are finite, while the
dissipation rate vanishes due to the interference of the density and current
excitations, cf. Eq.~(\ref{eq:momentum_dot}).

Finally we give some estimates for the critical force $F_c$. For $^{87}$Rb gas
with the linear density $n= 10^5\, {\rm cm}^{-1}$ and the 1D interaction
constant $g$ of the same order as $\hbar^2 n/m$, one estimates the 
sound velocity as  $c\approx 1.4\, \hbar\, n/m \approx 1\, {\rm cm/s}$. 
Taking $M=m$ and $M^*\approx 1.3\,m$ \cite{Fuchs_2005}, and the temperature 
$T\approx \, 0.5\, mc^2\approx 5\times 10^{-7}\, {\rm K}$, one finds for 
the {\em naive} critical force
$F_c^{(0)} \approx\pi^3 T^4/(c^4\hbar^2 n M^*)\approx 4 \times 10^{-22}\, {\rm
N}$. For comparison, the gravitational force acting on $^{87}$Rb atom is
$1.4\times 10^{-24}\, {\rm N}$, i.e. about $300$ times weaker.  However,
closeness to the integrability with say $G/g=1.03$ decreases the {\em actual}
critical force $F_c$ by about three orders of magnitude, see
Eqs.~(\ref{eq:kappa_weak}) or (\ref{eq:kappa_strong1}). This makes the
critical force to be well below the gravitational one, making it possible to
observe Bloch oscillations of the spin-flipped particle falling in the
gravitational field.

% \section{Acknowledgments}
% \label{sec:ack}

We are grateful to L.~I.~Glazman, M.~K\"ohl and K.~Bongs
for stimulating discussions.
D.M.G. acknowledges support by EPSRC Advanced Fellowship EP/D072514/1 and
a warm hospitality of
%V.E.~Zakharov
Mathematical Physics group
at Lebedev Physical Institute of Russian Academy of Sciences.
% and particularly the negligence of the guards paying no attention to
% the fake passes during  the manuscript preparation.
%Both AK and DMG are grateful
%to the International Telecom and Mobile Corporations for providing cheap
%communications indispensable for successful completion of this project across
%the Atlantic.
A.K.  was supported by NSF grants DMR-0405212,
DMR- 0804266 and acknowledges EPSRC grant GR/T23725/01 for the support of his stay in Birmingham.


\begin{thebibliography}{25}
\bibitem{foot1} According to  fluctuation-dissipation theorem,
the friction force is accompanied by  stochastic force $f(t)$ 
with the correlator
$\langle f(t)f(t')\rangle = 2\kappa T\delta(t-t')$.  The complete
semiclassical description of the particle kinetics is given by the
distribution function ${\cal P}(X,P)$, satisfying the Fokker-Planck
equation. Here we restrict ourselves with the dynamics of the maximum of this
distribution only.
\bibitem{foot3} There is also a cross-term
$-\Gamma_{\rho\vartheta} \rho(X) \derx\vartheta (X)$, where
$\Gamma_{\rho\vartheta} = V \left(1-\delta_l +\delta\!M/m \right)/n$, but
it does not contribute to the momentum relaxation rate.
\end{thebibliography}
\end{document}